\documentclass[letter]{aa}  
\usepackage{graphicx}
\usepackage{txfonts}

\begin{document}

\title{High-mass microquasars from binary to black hole scale}

\author{Rolf Walder\inst{1}
  \and
  Doris Folini\inst{1}}

\institute{Centre de Recherche Astrophysique de Lyon (CRAL), \'Ecole
  normale sup\'erieure de Lyon, Universit\'e de Lyon, Universit\'e
  Lyon 1, CNRS, UMR 5574, 46 all\'ee d’Italie, 69364 Lyon Cedex 07,
  France.  \email{Rolf.Walder@ens-lyon.fr.}}

\date{Received ; accepted }
\abstract{We present a 3D hydrodynamical simulation of a
  wind-accreting  high-mass microquasar, from 30 binary separations ($d$)
  to 256 black hole (BH) gravitational radii, over one-sixth of a full orbit
  in time, with system parameters inspired by Cyg~X-1. The simulation
  allows key system components to emerge naturally as inter-dependent
  quasi-stationary parts of an inherently multi-scale flow. The BH
  accretion disk is highly eccentric, with spirally shaped accreting
  and decreting zones. Its flow field is consistent with elliptical
  orbits confocal at the BH. The disk structure relates to its
  feeding: a cold 3D accretion cone channels matter from opposite the
  L1 point and within 2/3$d$ from the BH toward the disk. Above and
  below the disk, a polytropic atmosphere establishes, with
  temperatures one-tenth of the virial temperature. A hot cocoon of
  shocked wind material engulfs the BH accretion structure on scales
  of $d/10$. We hypothesize that the shocks may accelerate particles
  and the atmosphere may up-scatter photons to GeV energies and
  beyond. An Archimedian spiral is apparent out to at least 10$d$, as
  the orbiting BH perturbs the homogeneous donor star wind. Our
  simulation offers a coherent cross-scale perspective that allows us to
  contextualize observations, interpretations, and specific models.
}

\keywords{X-rays: binaries -- Accretion, accretion disk --
  Hydrodynamics -- Gamma rays: stars -- Acceleration of particles}
\maketitle

\section{Introduction}
\label{Sec_Intro}

Microquasars (MQs), where a black hole (BH) accretes mass from its
binary star companion, are among the most energetic stellar scale
phenomena.  MQs come in different flavors, depending on the donor
star, and have excellent observational coverage~\citep[for reviews
  see][]{2002RSPTA.360.1967D, 2006ARA&A..44...49R,
  2011MNRAS.411..337D}. We focus on MQs with a high-mass wind-shedding
donor, with Cyg~X-1 as the prototype system. We present the first 3D
hydrodynamical simulation on scales from circumbinary (30~binary
separations, $2\times10^{14}$~cm) to roughly BH ($10^8$~cm or 256
gravitational radii, $R_{\mathrm{G}} \equiv G M_{\mathrm{BH}}/c^2$).
The emerging self-consistent flow is rich in features: accretion disk,
accretion wake, hot atmosphere, bow shock, cocoon. The relevance of
our results is twofold. They highlight the intrinsically multi-scale
nature of the flow, and they offer a coherent picture to contextualize
the interpretation of the observations and the models designed to
scrutinize specific aspects of MQs.  This letter aims at a concise
presentation of key findings based on one simulation. Further
simulations will be published elsewhere.

\section{Physical and numerical model}
\label{Sec_PhysModel}

The system parameters used are from %inspired by
Cyg~X-1~\citep{2011ApJ...742...84O}\,: $M_{\mathrm{BH}} =
14.8$\,$M_{\sun}$, M$_{\mathrm{D}} =19.2$\,$M_{\sun}$, circular orbit,
5.6-day period, $R_{\mathrm{G}} = 21.85$~km, donor star radius
$R_{\mathrm{D}} = 16.2R_{\sun}$ \, ($10^{7}$~km),
binary separation $d=2.99 \times
10^{7}$~km. More recent estimates find 50\% to 100\% more massive
components in a wider system~\citep{2021Sci...371.1046M}. For the
donor we assumed an isotropic wind with mass loss $\dot{M}_{\mathrm{D}}=
10^{-6}$~M$_{\sun}$/y, wind speed $v_{\mathrm{D}} = 850$~km/s, and photospheric
temperature $T_{\mathrm{D}}=31'000$~K.

The physical model was kept simple. Ideal hydrodynamics were used with
Newtonian gravity, ideal equation of state with $\gamma = 5/3$, and
optically thin cooling (parameterized losses via lines and
recombination, online calculated losses via free-free and inverse
Compton of stellar UV photons~\citep{2022A&A...658A.100C}).  We relied
on numerical diffusion to model diffusive processes. Within the disk,
we prescribed a minimum temperature as a function of BH distance $r$,
inspired by~\citet{1973A&A....24..337S}\,: $T(r) = 10^9 \times
(R_{\mathrm{G}}/r )^{3/4}$~K. The BH was modeled as a spherical
accretor with radius $R_{\mathrm{Acc}}= 256$~$R_{\mathrm{G}}$, moving
on its orbit in the 3D grid (simulation frame). Adaptive mesh
refinement (AMR) ascertains sufficient resolution for all relevant
structures.  Starting at cell size $dx=dy=dz=10^{12}$~cm, 14 levels of
refinement (factor of two each) were used, with time steps from 90~s
globally to 51~ms near the BH (see details in
Appendix~\ref{App_NumericalModel}).

For post-processing, the AMR data were interpolated on a spherical grid
centered at and moving with the BH, spanning the sphere at $1\degr$
resolution in azimuthal ($0 \le \phi \le 360$) and polar ($0 \le
\theta \le 180$) direction, with 140 logarithmically equidistantly
spaced radial points out to $1.5 \times 10^5$~R$_{\mathrm{G}}$
($3.3\times10^{11}$~cm). The sign convention is that negative mass flux
means accretion onto the BH. Other important analysis tools are
movies (see Appendix~\ref{App_Movies}).

\section{Results}
\label{Sec_Results}

\begin{figure*}[tb]
  \centerline{
    \includegraphics[width=0.99\textwidth]{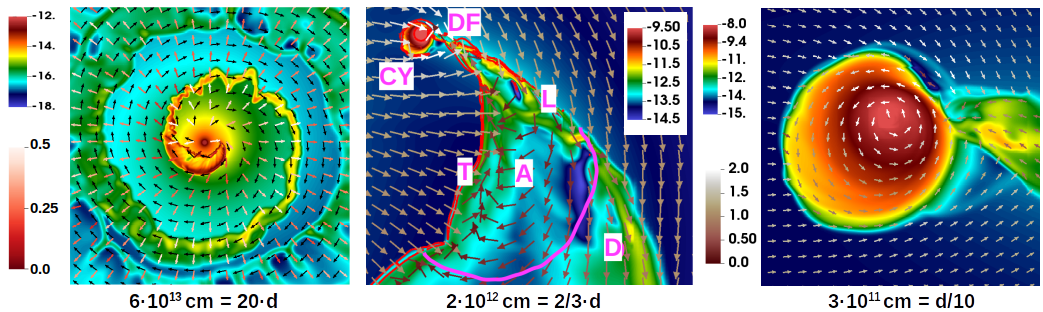}
    }
  \caption{Flow field in orbital plane, density [log10(g/cm$^{3}$)] and
    velocities [$10^{8}$~cm/s] from large ($20 \, d$) to small
    ($d/10$) scales. Left panel\,: BH-induced high-density spiral,
    aligned with the velocity field in the binary orbit co-rotating
    frame (black arrows; simulation frame velocities color-coded from red to
    white). Middle panel\,: BH accretion cone (A) with its
    leading (L) and trailing (T) edge, alimenting the BH courtyard
    (CY) via the disk-feeding region (DF). The boundary to the decretion
    zone (D) is in pink, strong shocks 
    in red. Right panel\,: Zoomed-in image on 
    accretion disk and feeding region. The velocities in the middle and
    right panels are in the frame co-orbiting with the BH (color-coded from brown to
    white).}
\label{Fig:Figure1}
\end{figure*}

\subsection{Wind-structure: Accretion and decretion cone}
\label{SubSec_Results_Cones}

The orbiting BH structures the flow field from accretor scale to
beyond binary separation (Fig.~\ref{Fig:Figure1}). Out to at least
10$d$, the wake of the moving BH imprints on the homogeneous donor
star wind as a high-density spatially structured Archimedian spiral
(Fig.~\ref{Fig:Figure1}, left, and Movie~1,
Sect.~\ref{Sect:Movie1}). In the reference frame co-rotating with the
binary components, the flow velocities align with the spiral. In the
simulation frame, they point radially away from the donor and are
modulated by the spiral. With increasing distance from the BH,
the wake widens and the neighboring windings of the spiral blur. The
circumbinary wind structure potentially matters for the
interpretation of observations and jet dynamics. Dense clumps in the
flow may lead to X-ray dips lasting minutes to
hours~\citep{2000MNRAS.311..861B, 2000A&A...354.1014D}.

The 3D accretion cone, where matter within the wake really accretes
onto the BH, extends out to $\sim 2/3 d$ from the BH (Fig. 1, middle,
and Movie~2, Sect.~\ref{Sect:Movie1}). The flow in the cone is cold,
yet ionized, potentially emitting in the UV. Exact temperature
estimates would require detailed radiative transfer. The confining
leading (L) and trailing (T) edge differ fundamentally. The former
moves into the donor star wind; the latter gets pushed by the wind. At
T the wind inclination angle is steeper, the post-shock temperatures are
higher, and the velocities directed away from the BH are smaller. This is
why the dividing surface between the accretion and decretion zone is
tilted (pink line). Instabilities are seen primarily at L. Their
analysis is beyond the scope of this letter. Post-shock temperatures
at either L or T reach $\sim$$10^{6}$K, implying X-ray and extreme ultraviolet (EUV) emission
from the narrow cooling layers.  Farther away, the shocks 
turn into discontinuities that separate the wake from the unperturbed
wind.

At distances of~$\sim d/10$ around the BH emerges what we term the
BH courtyard (CY, Fig.~\ref{Fig:Figure1}, right): the richly
structured, inherently 3D, near-BH flow field. The CY, detailed below,
comprises a moderately hot cocoon of shocked donor star wind, within
which the accretion disk is embedded, along with a much hotter 
atmosphere and the disk feeding region (DF), where the 3D accretion
cone attaches to the disk.

\subsection{BH courtyard\,: Cocoon and feeding of disk}
\label{SubSec_Results_CY_Feeding}

\begin{figure*}[tb]
  \centerline{
    \includegraphics[width=0.99\textwidth]{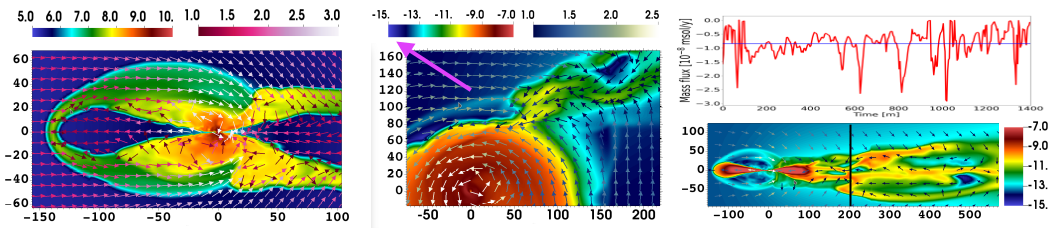}
    }
  \caption{BH courtyard. Left panel\,: Temperature [log10(K)] and
    velocities (arrows, white to red) in a slice normal to the orbital plane
    and through BH (at $x=0$, $y=0$) and accretion cone.  Visible are
    the cocoon (bow shock flow, green, on the left), the atmosphere
    (red and yellow, above and below the BH), the disk (dark blue),
    and the accretion cone (dark blue) with engulfing shock (green and
    yellow, on the right). Middle panel\,: Disk feeding region in
    orbital plane, density [log10(g/cm$^{3}$)] and velocities (arrows,
    blue to white). The BH moves in the direction of the pink
    arrow. Right panel, top\,: Time-dependent mass flow to the BH (in
    $10^{-8}$~M$_{\sun}$/y, time mean in blue), integrated over the
    accretion cone at $2\times10^{11}$~cm from the BH (black line in
    bottom right panel).  Right panel, bottom\,: Density
    [log10(g/cm$^{3}$] and velocities (same coloring as middle panel,
    same slicing plane as left panel). Lengths are in units of
    $10^{9}$\,cm. Velocities are in the frame co-orbiting with the BH, projected onto
    the slice shown, in $10^{8}$\,cm/s.}
\label{Fig:DISK_Inlet}
\end{figure*}

The outermost structure of the BH CY is a moderately hot cocoon, which
occupies a volume of a few $10^{33}$~cm$^3$ and arises from the bow
shock caused by the orbiting BH with its disk. With temperatures in a
range $10^{6.5} < T < 10^{7.5}$~K, X-ray and EUV emission can be
expected (Fig.~\ref{Fig:DISK_Inlet}, left panel). The flow is smooth
and roughly parallel to the orbital plane, until it interacts with the
hot envelope engulfing the accretion cone, upon which temperatures rise
up to $T \approx 10^{8.5}$K. There is no material falling directly
from the cocoon into the BH.

Fresh matter reaches the disk via the 3D accretion cone. Identifying
the cone via deviations from the donor star wind density, we find half
opening angles of $24\degr \pm 8\degr$ and $16\degr \pm 5\degr$ in the
azimuthal and polar direction, respectively (time mean and standard
deviation). At a distance of $2\times 10^{11}$ cm from the BH, the
extent of the cone in polar direction is comparable to that of the
cocoon (Fig.~\ref{Fig:DISK_Inlet}, bottom right). The matter within
the cone is highly structured and cold, except for thin shock-heated
layers wrapping the cone. The within-cone mass flow toward the BH is
time variable, with peaks exceeding the mean by a factor of four
(Fig.~\ref{Fig:DISK_Inlet}, top right, and Movie~2,
Sect.~\ref{Sect:Movie2}). Occasionally, streamers from the accretion
cone bypass the disk entirely, entering the BH directly from above or
below (see Movie~2, Sect.~\ref{Sect:Movie2}). The variable feeding may
contribute to observed variable emissions or time lags between UV and
X-ray~\citep{2013MNRAS.434.1476I, 2024MNRAS.530.4850H,
  2025MNRAS.536.3284U}.

The accretion cone attaches to the disk in the disk feeding
region, located roughly opposite  the L1 point
(Fig.~\ref{Fig:DISK_Inlet}, middle and bottom right). At disk entry,
the cone is oriented neither along a straight line to the BH nor
tangentially to the disk (Fig.~\ref{Fig:DISK_Inlet}, middle panel, and
Movie~3 Sect.~\ref{Sect:Movie3}). This permanent perturbation to the
disk gives rise to a quasi-stationary, non-Keplerian flow within the
disk (see~\ref{SubSec_Results_Disk}). The location and cone
orientation are set by the balance between gravity and stellar wind
momentum entering the cone through its leading and trailing edges.

\subsection{BH courtyard\,: Atmosphere}
\label{SubSec_Result_BH-atmosphere}

\begin{figure}[tb]
  \centerline{
    \includegraphics[width=0.53\textwidth]{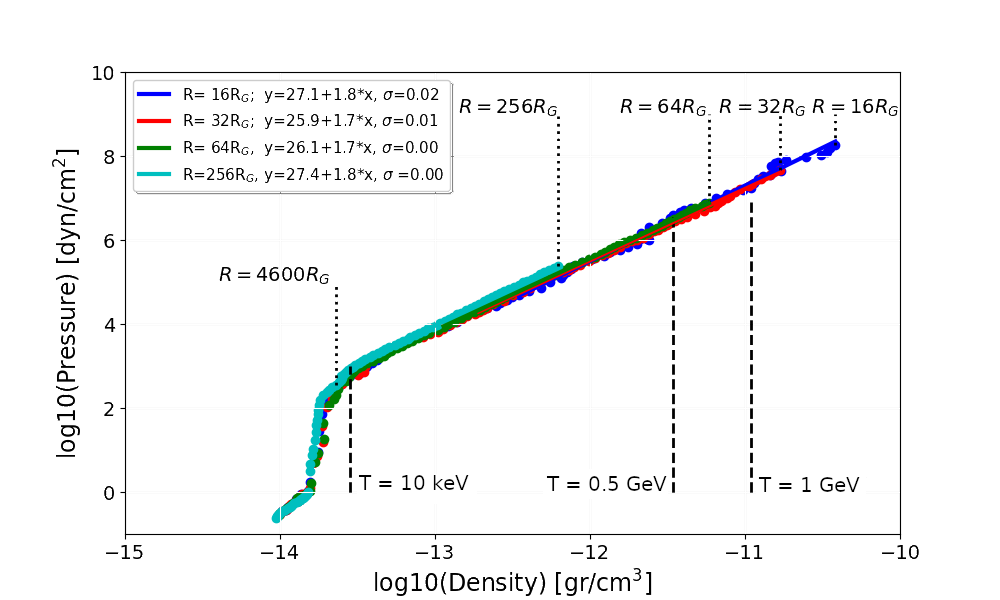}
   }
  \caption{Pressure-density-polytrope of the BH atmosphere. Shown are
    the data (dots) from the simulations with different sizes of the accreting
    sphere and their fits (solid lines);
    $R_{\mathrm{Acc}}=256$~R$_{\mathrm{G}}$ (cyan),
    $R_{\mathrm{Acc}}=64$~R$_{\mathrm{G}}$ (green),
    $R_{\mathrm{Acc}}=32$~R$_{\mathrm{G}}$ (red),
    $R_{\mathrm{Acc}}=16$~R$_{\mathrm{G}}$ (blue).}
\label{Fig:BH_Atmosphere}
\end{figure}

A polytropic quasi-static atmosphere develops above and below the BH
and its accretion disk. Embedded in the cocoon, the atmosphere
distinguishes itself by a less ordered flow field and hotter
temperatures (Fig.~\ref{Fig:DISK_Inlet}, left panel). Locally,
velocities reach a few $10^{9}$~cm/s and point in any direction. The
median flow is toward the BH; it is subrelativistic and 
subsonic, and contributes less than 0.1\% to the accretion rate. The
density is inhomogeneous and intermittent in time. A tiny part of the
density probability distribution function typically reaches
values one to two orders of magnitude above the median, implying
strongly different mean and median density (not shown). Median values
of pressure and density fit a polytrope with a slightly superadiabatic
index $1.7 \le \gamma \le 1.8$ (Fig.~\ref{Fig:BH_Atmosphere}). The
finding is robust against concurrently shrinking the
accretor and increasing the refinement (see
Appendix~\ref{App_NumericalModel}). Temperatures are $\sim
10$\% of the virial temperature. The polytrope implies that electrons
get relativistic at a distance of $\sim 100 R_{\mathrm{G}}$ and
that pair production becomes energetically possible at $\sim 45
R_{\mathrm{G}}$.  The atmosphere is alimented by shock-heated matter
from around the accretion cone, the cocoon, and the internal shock
forming between the two flows (not shown).

\subsection{BH courtyard\,: Accretion disk}
\label{SubSec_Results_Disk}

\begin{figure*}[tb]
  \centerline{
    \includegraphics[width=0.99\textwidth]{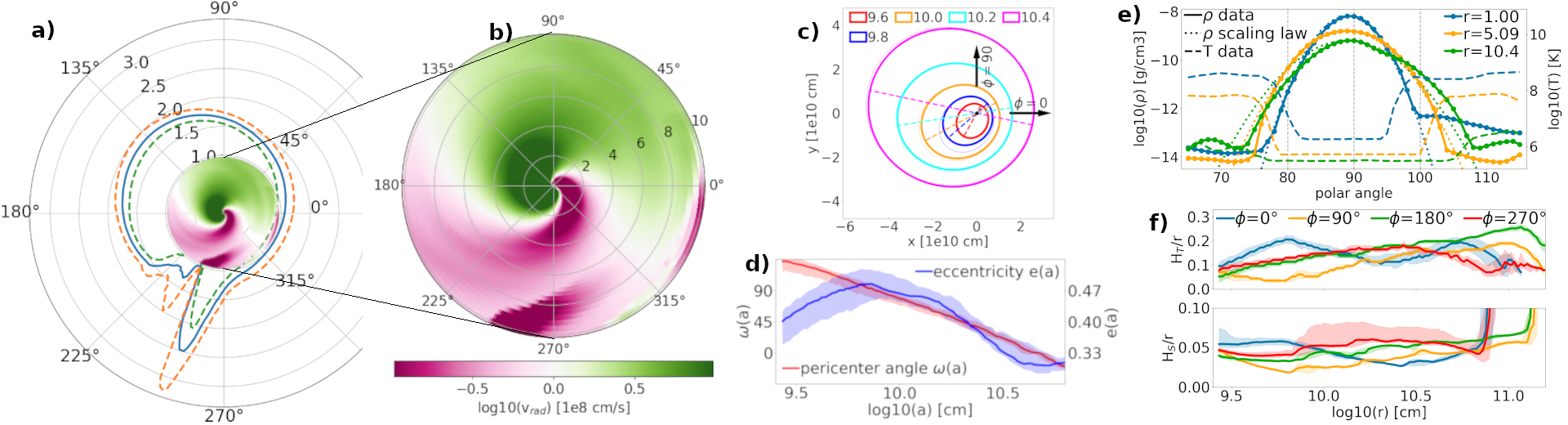}
    }
  \caption{Accretion disk characteristics. Quasi-stationary extent of
    disk (contours, radius in $10^{11}$cm, panel a) and
    radial velocity in orbital plane (in BH co-orbiting frame, radius in
    $10^{10}$cm, panel b), inferred elliptical orbits (for one time
    and five color-coded pericenter distances log10($r$),
    panel c), eccentricity and pericenter angle vs. semimajor axis
    (panel d), density and temperature profiles (for one time,
    $\phi=180\degr$), and three radii (in $10^{10}$cm, colors), from data (solid and
    dashed) and via scaling law (dotted, panel e), disk thickness
    $H_{\mathrm{T}}/r$ and scale height $H_{\mathrm{S}}/r$ (panel f). The dashed lines
    (panel a) or shading (panels d and f) indicate the
    25${\mathrm{th}}$ to 75${\mathrm{th}}$ percentile with respect
    to time.  The circles in panel e indicate the $1\degr$ rays.}
\label{Fig:DISK}
\end{figure*}

Around the BH, an eccentric quasi-stationary disk forms (see
Fig.~\ref{Fig:DISK}). In the orbital plane, the radial extent of the
disk varies by about a factor of two as a function of $\phi$
(Fig.~\ref{Fig:DISK}a).  It is smallest between the disk feeding
region ($\phi \approx 260$) and the direction of orbital motion
($\phi=0$), and largest in the opposite quadrant ($90 \le \phi \le
180$).  Temporal variability is comparatively small (dashed
25${\mathrm{th}}$ and 75${\mathrm{th}}$ percentile contours). The
within-plane flow field takes Keplerian values $v_{\mathrm{K}}(r)$, if
averaged over $\phi$. As a function of $\phi$ and BH distance $r$, the
azimuthal velocity component $v_{\mathrm{\phi}}(r,\phi)$ deviates by
about $\pm 20$\% from $v_{\mathrm{K}}(r)$. Radial velocities
$v_{\mathrm{r}}$ are organized in a double-spiral pattern of inward
and outward directed flow (in pink and green in Fig.~\ref{Fig:DISK}b
and Movie~3, Sect.~\ref{App_Movies}). The accretion cone is visible
around $\phi \approx 260\degr$ for $r>6 \times 10^{10}$cm.  The flow
field is in line with a set of confocal elliptical orbits, illustrated
for a snapshot in time in Fig.~\ref{Fig:DISK}c. The BH is the common
focus; eccentricity $e(a)$ and pericenter phase $\omega(a)$ vary with
semimajor axis $a$ (Fig.~\ref{Fig:DISK}d). The underlying analysis
assumed that $v_{\mathrm{\phi}}(r,\phi)$ assumes its maximum
$v_{\mathrm{\phi}}^{\mathrm{max}}$ over $\phi$ for fixed $r$ at
pericenter, implying pericenter angle $\omega=\phi_{\mathrm{peri}}$,
eccentricity $e=(v_{\mathrm{\phi}}^{\mathrm{max}}/ v_{\mathrm{K}})^{2}
-1$, and semimajor axis $a=r/(1-e)$.

Orthogonal to the orbital plane, in a wedge with half opening
angle of $\sim10\degr$, density decreases by four to six orders of
magnitude, depending on radius (Fig.~\ref{Fig:DISK}e, solid
lines). Temperatures (dashed lines) are isothermal at the imposed
minimum value (see Sect.~\ref{Sec_PhysModel}) within a narrower
wedge. Simulated density profiles are compatible with isothermal
profiles $\rho(z) = \rho_{\mathrm{0}} \times
\exp(-z^{2}/2H_{\mathrm{S}}^{2})$, where $\rho_{\mathrm{0}}$ is the
central density (dotted lines). We infered the local scale height
$H_{\mathrm{S}}(r,\phi,t)$  from our data via identifying
$z(r,\phi,t)$ such that $\rho(z) = \rho_{\mathrm{0}}/1000$. Systematic
dependences  of $H_{\mathrm{S}}$ on $\phi$ and $r$ can reach a factor
of two, dominating over the temporal variability of $H_{\mathrm{S}}$
(Fig.~\ref{Fig:DISK}f). Values of $H_{\mathrm{S}}$ mostly fall within
$\pm50$\% of the scale height of an isothermal disk,
$H_{\mathrm{S,iso}}(r,\phi) = c_{\mathrm{s}}/v_{\mathrm{K}}$ (not
shown). In the inner parts of the disk, at $r=10^{10}$cm,
$H_{\mathrm{S}} \approx 0.05 \times r \approx 5\times10^{8}$cm or
about four AMR grid cells, which may be considered a limitation of the
presented results.

The disk geometry and in-plane velocity field are in qualitative
agreement with analytical work by~\citet{2024A&A...686A.264R}. The
permanent perturbation required in their model is provided by the
quasi-stationary location and orientation of the disk feeding.  Our results
further demonstrate that the disk has a finite, $\phi$-dependent
vertical extent. The latter is also reflected when examining radial
mass fluxes. Staying in the orbital plane ($\theta=90$), regions of
negative (toward the BH) and positive radial mass flux combine into a
net positive radial mass flux. The net mass flux turns negative and
reaches approximately the BH accretion rate only when integrated over a wedge
of roughly $\pm 10\degr$ in polar direction (not shown). This is in
line with a model presented by~\citet{2018MNRAS.479.1528B}.

\section{Discussion and conclusions}
\label{Sec_Discussion}

Although simple in terms of physics, our 1400 minutes of 3D simulations
from system scale to accretor scale demonstrate the emergence of a
richly structured flow, which is essentially 3D on all scales and where
cross-scale interactions play a key role.

On scales of about 1/10 the system separation, a moderately hot cocoon
establishes behind the bow shock that arises as the BH with its
accretion disk moves through the donor star wind. Embedded within the
cocoon are a polytropic atmosphere and the quasi-stationary,
eccentric, BH accretion disk. The atmosphere is alimented via
collision of cocoon matter with shock-heated matter wrapping the
accretion cone, resulting in temperatures $T \geq 10^8$~K. The disk
height-to-radius ratio of $0.05 \le H/r \le 0.1$ varies with azimuthal
angle, in line with the in-disk velocity field following elliptical
orbits whose eccentricity varies with the pericenter distance to the
BH. The disk structure is tightly linked to  the position of the
disk feeding region (opposite the L1 point) and to the orientation of the
accretion cone (neither tangential to the disk nor toward the
BH). These, in turn, are set by the interplay between donor star wind
and BH gravity.  We therefore expect some sensitivity of our results
to the donor star wind, including any anisotropy or clumpiness, and
the (potentially non-spherical) binary orbit.
  
Despite the simplistic model, our results invite speculations on
potential implications. Cyg~X-1, which inspired this study, shows
persistent non-thermal emission into the MeV range, and transient
emissions in the GeV and TeV range~\citep[][]{2016A&A...596A..55Z,
  2024arXiv241008988L}. Various emission mechanisms have been
suggested. They typically rely on photon up-scattering and particle
acceleration for hard tails~\citep[][]{2025MNRAS.540..205A}. Different
system components have been put forward that offer favorable
conditions for these mechanisms to operate efficiently.

Our simulations naturally develop conditions where these mechanism
may efficiently take place. The hot portion of the BH courtyard
(cocoon, atmosphere) offers electrons to up-scatter X-ray or
UV photons. Shocks as potential sites of particle acceleration exist,
within the BH courtyard or confining it. It is plausible that
the flow in this region is collisionless: the ratio of the
electron mean free path to their inertial length is about 10'000; the
Coulomb logarithm is around 7. Particle acceleration may be favored by
double-shock structures~\citep{2023PhRvE.107b5201M} and corrugated
shocks~\citep{Demidem_2023}, both found in our
simulations. Admittedly, two system components often invoked for
emissions are not covered by our simulations: jets and the vicinity of
the BH on some ten $R_{\mathrm{G}}$.

The presented model leaves ample room for improvement. We
consider three axes of improvement to be the most crucial. First,  to better
capture the thermal structure, some form of radiative transfer should
be included, as well as two temperature flows to separately account
for electrons and ions. Second, magnetic fields should be included,
associated with the disk and with the donor star. Supergiants
carry a magnetic field with a large variety of field topologies and
strengths \citep{2012SSRv..166..145W, 2019MNRAS.489.5669P}.  In a
recent review, \citet{2022hxga.book...46U} estimate that 10\% of all
O- and B-stars harbor strong globally ordered (mostly bipolar)
magnetic fields. These fields can have a significant impact on
the formation and isotropy of the stellar wind. Even weak and
unordered fields provide a seed for a possible dynamo process in
the inner disk. A strong field may result, which again impacts
  the disk, notably its angular momentum transport, and the launching
  and stability of any jet. Third, the central engine of the MQ must be
resolved and physically addressed more comprehensively, including jet
launching. The presented simulations treat the BH as an unresolved
accretor with $R_{\mathrm{Acc}} \leq 256 R_{\mathrm{G}}$.

We advocate that the presented results are nevertheless of interest:
as a baseline against which follow-up studies addressing additional
physics or a wider portion of the parameter space may be compared; as
a proof of concept that the central BH engine can be studied in its
large-scale context using AMR; and as a self-consistent global scale
picture, which may assist the interpretation of observations and the
design of dedicated studies focusing on particular aspects of MQs.

\begin{acknowledgements}
  This work was performed using HPC resources from GENCI (Grant
  A0150406960) and using ressources of CBPsmn at ENS Lyon.
\end{acknowledgements}

\bibliographystyle{aa}
\bibliography{MultiScaleMQ}

\begin{appendix}

\section{Numerical model of the simulation}
\label{App_NumericalModel}

The simulations were performed with the numerical toolbox A-MaZe
\citep{2000ASPC..204..281W, 2019A&A...630A.129P,
  2022A&A...658A.100C}. Cartesian meshes in a fixed Eulerian frame
(simulation frame) were used. Fluxes through cell-interfaces were
computed with an exact Riemann-solver. Minmod limiters provided
sufficient diffusion at strong gradients, notably shocks. For
integration in time we used an unsplit strong stability preserving
(SSP) Runge-Kutta method of second order (for details, see 
\citet{2019A&A...630A.129P}). Other grid choices have been
  explored by the authors to some degree over the years (e.g., cubed
  spheres, skewed grids, or mapped grids), and were found to have 
  their pros and cons when it comes to simulating binary stars
  including orbital motion in 3D.

The A-MaZe toolbox features two AMR algorithms. Here,
block-structured AMR was used, as detailed
in~\citet{2008A&A...484L...9W}. Finer and finer meshes are constructed
around the accreting sphere, using a refinement ratio of two between
successive levels of refinement over a total of 14 levels. The
algorithm refines not only space, but also time, with the same
refinement ratio across levels. For one coarse level time step, the
next finer level is integrated twice in time, with half the coarse
level time step. In this way, the same cfl condition can be used on
all refinement levels. Here, a cfl of 0.5 is used. For each level of
refinement, the algorithm subdivides the entire refined region into
approximately equally sized grids. The number of grids corresponds to
the number of processors used, such as to ascertain good load
balancing. The coarsest level (level 1) extends in each
direction over $2\times10^{14}$~cm and has a grid spacing of
$dx=dy=dz=10^{12}$~cm. Grid spacing on level 14, the finest level
  of refinement, is $dx=dy=dz=1.2\times10^{8}$~cm. Levels 1 to 5 are
  kept fixed in space and time. Level 5 has a spatial extent of
  $6.4\times10^{12}$cm and is the last (finest) level to still cover
  the entire binary orbit. Refinement levels 6 to 14 have fixed
  spatial extent, but their position in the computational domain is
  time dependent. These levels follow the orbital motion of the
  accreting BH. The AMR algorithm achieves this in the following
  way. The levels (grids) are kept fixed in space during 4 integration
  steps on level $l$. Then, new grids are constructed on levels $l$ to
  14, such that these new levels are again centered at the
  accretor. The new grids on each individual level are initialized
  using data from the old grids of this level where available, i.e.,
  at locations where the new and the previous position of the level
  overlap. Where this is not the case, data from the next coarser
  level is used.

The donor star as well as the BH moved within the computational box on
circular, Keplerian orbits. The spherical donor star was covered with
about $10^{6}$ grid cells (refinement level 7). Mass was shed
isotropically in the reference frame of the donor star. Numerically,
this was achieved by mapping corresponding densities, temperatures, and
(supersonic) velocities onto all grid cells within the donor star,
before integration the next time-step. The accreting sphere (BH) was
covered by about 500 grid cells and was modeled by what is often called
an ``all absorbing condition.'' In this approach, which is widely used,
all material entering the sphere is removed from the mesh after each
individual fine grid time step. As shown
in~\citet{1995A&A...301..922Z}, this is to a high degree equivalent to a free
flow boundary condition with elaborate spherical reconstruction of the
sphere's surface. In particular, as shown
in~\citet{1995A&A...301..922Z}, the 'all absorbing condition' preserve
the correct stability condition and does not produce spurious
oscillations around the accretor.

The simulation was done in three steps. The goal of the first
step is to reach a quasi-relaxed large-scale circumbinary
structure. Only six levels of refinement were used, the numerical radius
of the accreting sphere (the BH) was large ($R_{\mathrm{Acc}} = 65'536
R_{\mathrm{G}}$). The simulation can then be advanced for a longer
time. The large-scale, circumbinary structure can settle and the
accretion flow is correctly embedded into the binary structure. The
goal of the second step is to increase the resolution around the BH,
so  as to enable the formation of an accretion disk and the BH
courtyard (see main body of paper). To this end, the radius of the
accreting sphere was successively reduced by factors of two, each time
adding an additional level of refinement to guarantee sufficient
numerical resolution around the accreting sphere. After each
additional refinement, the flow structure was again relaxed to a
quasi-stationary state (ascertained via inspection by eye). In this
way, large and small scales always match. Refinement continues until
we reach 14 levels of refinement ($dx=1.2\times10^{8}$~cm) and an
accreting sphere of $R_{\mathrm{Acc}} = 256 R_{\mathrm{G}}$ (about
$5.6\times10^{8}$~cm). For $R_{\mathrm{Acc}} \gtrsim 1000
R_{\mathrm{G}}$, the bow shock around the BH remains attached to the
accreting sphere. For smaller radii, it detaches and moves in front of
the BH -- a decisive step to take since the flow characteristics and
stability properties drastically change at this transition
\citep{1995A&A...301..922Z}. For $R_{\mathrm{Acc}} = 516
R_{\mathrm{G}}$, a disk-like structure forms, which develops into a
proper disk once $R_{\mathrm{Acc}} = 256 R_{\mathrm{G}}$. The goal of
the third step is to collect the 1400 minutes (about one-sixth of a full
orbit) of data presented in this paper, representing the
quasi-stationary situation with all relevant structures of the flow\,:
cocoon, BH atmosphere, disk feeding region, and disk. This step takes
longer than the first two steps. The structure first needs to relax
again (as in steps one and two, these data are finally discarded),
then the actual data collection can take place. It is this last part,
the actual data collection, which consumes most of the total compute
resources.

The computational costs are considerable: using 16 cores, the
acquisition of the 1400 minutes of data took three months. The costs
are essentially set by the resolution around the accretor. Our
run-statistics indicate that the two to four finest levels of
refinement -- covering the vicinity of the accretor -- account for
about 75\% of the total computational costs, intermediate scales (up
to 2 times the binary separation) for about 20\%, while the large
scale demand less than 5\%. The costs to establish the reliable
quasi-stationary structures are negligible, compared to the costs
needed to cover one-sixth of the orbit when advancing the quasi-stationary
state.

Simulations with even smaller $R_{\mathrm{Acc}}$ reveal even more
features on scales close to the accretor, but only shorter time
periods can be simulated because of computational costs. For
Fig.~\ref{Fig:BH_Atmosphere}, we used data of simulations with
smaller accreting spheres that run on 64 cores, notably\,:
a)~$R_{\mathrm{Acc}}= 128$~R$_{\mathrm{G}}$ with a simulation time
$T_{\mathrm{S}}= 4$~h and simulation time step on the finest level,
the scale of the accretor of $Dt_{\mathrm{f}} \approx 1.8$~ms;
b)~$R_{\mathrm{Acc}}= 64$~R$_{\mathrm{G}}$ ($T_{\mathrm{S}}= 70$~min,
$Dt_{\mathrm{f}}\approx 0.45$~ ms); c)~$R_{\mathrm{Acc}}=
32$~R$_{\mathrm{G}}$ ($T_{\mathrm{S}} = 12$~min,
$Dt_{\mathrm{f}}\approx 0.14$~ms); d)~$R_{\mathrm{Acc}}=
16$~R$_{\mathrm{G}}$ ($T_{\mathrm{S}}= 2.5$~min,
$Dt_{\mathrm{f}}\approx 0.07$~ms). As can be seen, $Dt_{\mathrm{f}}$
decreases typically even by more than a factor of two for each
additional level of refinement, making the computational cost issue
even more severe. The reason is particular to the physical problem
under consideration, a bottleneck due to physics and not
  particular to the AMR approach used here. As $R_{\mathrm{Acc}}$
shrinks, velocities in the vicinity of the accreting sphere
increase. To respect the cfl condition, the time step on the finest
level, near the accretor, must be reduced. The reduced time step then
propagates up to the coarsest level. These additional simulations thus
are not advanced as far in time as our main simulation
($R_{\mathrm{Acc}}= 256$~R$_{\mathrm{G}}$) and have not as good
temporal statistics. Nevertheless, from Fig.~\ref{Fig:BH_Atmosphere}
it can be taken that a main finding - polytropic atmosphere with $1.7
\le \gamma \le 1.8$ - is robust against reducing the size of the
accretor, only the range over which the polytrope extends
increases. Further, preliminary analysis of the results of these more
finely resolved models -- in terms of accretion rate, the shape, size,
height, and eccentricity of the disk, and the polytrope of the BH
atmosphere -- shows that results overall compare well with the main
model that is coarser but is further advanced in time and thus has
(much) better temporal statistics.

For those parts of the analysis where BH centered quantities are examined,
notably in Sect.~\ref{SubSec_Results_CY_Feeding},
Sect.~\ref{SubSec_Result_BH-atmosphere}, and
Sect.~\ref{SubSec_Results_Disk}, we relied on the AMR data remapped on a
spherical grid of equidistant rays, as described in
Sect.~\ref{Sec_PhysModel}. We tested that analyzed quantities are robust against
using a finer grid with rays separated by $0.5\degr$ instead of $1\degr$. 

For data analysis and visualization we used python and
VisIt\footnote{https://visit-dav.github.io/visit-website/index.html}.

\section{Movies}
\label{App_Movies}

As additional material, we provide four movies, showing the simulation
of 1400~minutes of the system~Cyg-X1 presented in this letter. As many
of the results presented concern stability, movies are an appropriate
medium to further deepen the insight into the governing physical
processes and the inter-dependence of scales.

\begin{figure}[tb]
  \centerline{
    \includegraphics[width=0.5\textwidth]{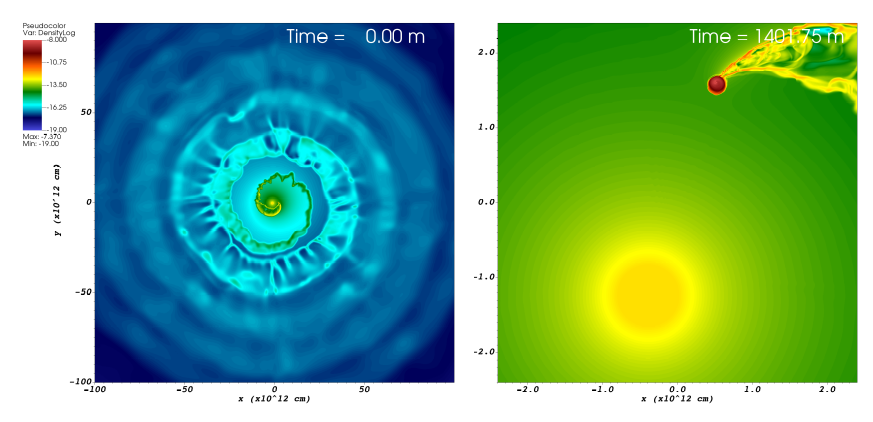}
    }
  \caption{Movie 1. Left panel\,: Density within the orbital plane, 2D
    extract of the full 3D computational domain. Right panel\,:
    Density within the orbital plane, 2D extract of the full 3D data
    on the orbit scale. We note the position of the BH at time 0.00~m and
    the position at time 1401.75~m.}
\label{Fig:App_Movie_Global}
\end{figure}

\begin{figure}[tb]
  \centerline{
    \includegraphics[width=0.5\textwidth]{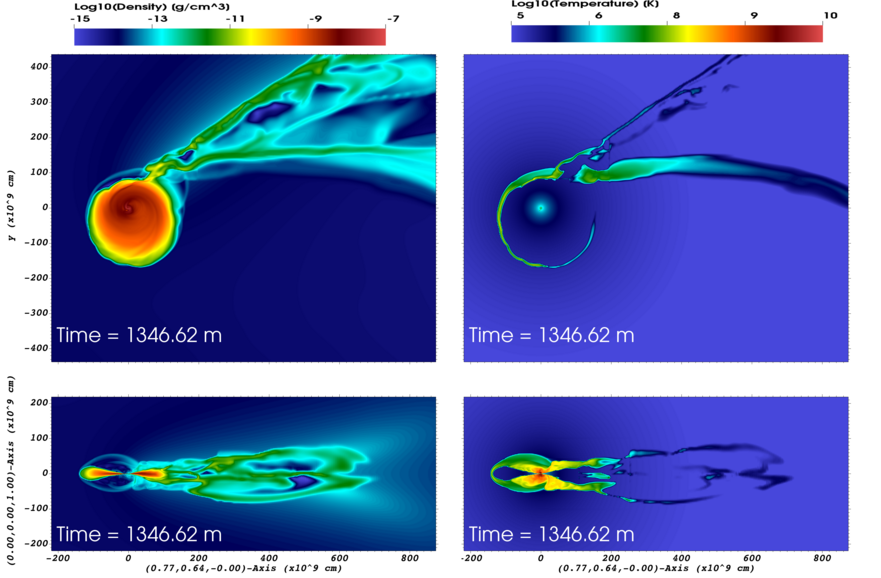}
    }
  \caption{Movie 2. Upper row: Density (left panel) and
    Temperature (right panel), 2D extract of the full 3D data on a
    scale concentrating on the region where the accretion stream feeds
    the accretion disk. Lower row: Density and temperature in
    a plane normal to the orbital plane. We note that this normal plane
    is always oriented within the accretion stream, thus turning
    during the simulation.}
\label{Fig:App_Movie_Disk_Inlet}
\end{figure}

\subsection{Movie 1\,: Movie1\_Scales.mp4}
\label{Sect:Movie1}

This movie aims to illustrate the simulation in general. Shown is
density in the orbital plane. At the beginning, the entire
computational domain ($2\times 10^{14}$~cm) is shown at the time where
the 1400 minutes of data taking start (see~\ref{Fig:App_Movie_Global},
left).  The movie then zooms into the binary star system
$5\times 10^{12}$~cm $= 5/3 d$ (see Fig.~\ref{Fig:App_Movie_Global},
right). Now the spatial view in the simulation frame is kept fixed and
the binary components along with the flow field advance in time for
1400 minutes, corresponding to about 1/6 of a full orbit.

\subsection{Movie 2\,: Movie2\_DiskInlet.mp4}
\label{Sect:Movie2}

This movie (see also Fig.~\ref{Fig:App_Movie_Disk_Inlet}) illustrates the
unsteady feeding of the accretion disk in density (on the left) and
temperature (on the right) in two different slicing planes. Shown in
the top row is the orbital plane. The bottom row shows a slice in a
plane normal to the orbital plane and cutting through the accretion
stream. Streamers bypassing the disk and hitting the BH directly
occur, for example, around minutes 641, 685, 728, 767, 842, 1004,
1220, 1307.

\subsection{Movie 3\,: Movie3\_DiskSpirals.mp4}
\label{Sect:Movie3}

This is the animation of Fig.~\ref{Fig:DISK}, panel b). The spirally
shaped pattern of radial velocities shows some disturbances at times
where the disk is fed by strong streams, compare with
Fig.~\ref{Fig:DISK_Inlet}, lower right panel.

\subsection{Movie 4\,: Movie4\_DiskEllipses.mp4}
\label{Sect:Movie4}

This is the animation of Fig.~\ref{Fig:DISK}, panel c). Remarkably,
ellipses and pericenter remain nearly stable over the simulation time.

\end{appendix}

\end{document}